\documentclass[11pt]{article}
\usepackage{fullpage}
\usepackage{url}
\usepackage{amsmath}

\begin{document}

\newcounter{savenumi}
\newenvironment{savenumerate}{\begin{enumerate}
\setcounter{enumi}{\value{savenumi}}}{\end{enumerate}
\setcounter{savenumi}{\value{enumi}}}
\newtheorem{theoremfoo}{Theorem}[section] %by chapter in report style
\newenvironment{theorem}{\pagebreak[1]\begin{theoremfoo}}{\end{theoremfoo}}
\newenvironment{repeatedtheorem}[1]{\vskip 6pt
\noindent
{\bf Theorem #1}\ \em
}{}

\newtheorem{lemmafoo}[theoremfoo]{Lemma}
\newenvironment{lemma}{\pagebreak[1]\begin{lemmafoo}}{\end{lemmafoo}}
\newtheorem{conjecturefoo}[theoremfoo]{Conjecture}
\newtheorem{research}[theoremfoo]{Line of Research}
\newenvironment{conjecture}{\pagebreak[1]\begin{conjecturefoo}}{\end{conjecturefoo}}

\newtheorem{conventionfoo}[theoremfoo]{Convention}
\newenvironment{convention}{\pagebreak[1]\begin{conventionfoo}\rm}{\end{conventionfoo}}

\newtheorem{porismfoo}[theoremfoo]{Porism}
\newenvironment{porism}{\pagebreak[1]\begin{porismfoo}\rm}{\end{porismfoo}}

\newtheorem{corollaryfoo}[theoremfoo]{Corollary}
\newenvironment{corollary}{\pagebreak[1]\begin{corollaryfoo}}{\end{corollaryfoo}}

\newtheorem{claimfoo}[theoremfoo]{Claim}
\newenvironment{claim}{\pagebreak[1]\begin{claimfoo}}{\end{claimfoo}}

\newtheorem{openfoo}[theoremfoo]{Open Problem}
\newenvironment{open}{\pagebreak[1]\begin{openfoo}\rm}{\end{openfoo}}

\newtheorem{exercisefoo}{Exercise}
\newenvironment{exercise}{\pagebreak[1]\begin{exercisefoo}\rm}{\end{exercisefoo}}

\newcommand{\fig}[1] %usage:\fig{file}
{
 \begin{figure}
 \begin{center}
 \input{#1}
 \end{center}
 \end{figure}
}

\newtheorem{potanafoo}[theoremfoo]{Potential Analogue}
\newenvironment{potana}{\pagebreak[1]\begin{potanafoo}\rm}{\end{potanafoo}}

\newtheorem{notefoo}[theoremfoo]{Note}
\newenvironment{note}{\pagebreak[1]\begin{notefoo}\rm}{\end{notefoo}}

\newtheorem{notabenefoo}[theoremfoo]{Nota Bene}
\newenvironment{notabene}{\pagebreak[1]\begin{notabenefoo}\rm}{\end{notabenefoo}}

\newtheorem{nttn}[theoremfoo]{Notation}
\newenvironment{notation}{\pagebreak[1]\begin{nttn}\rm}{\end{nttn}}

\newtheorem{empttn}[theoremfoo]{Empirical Note}
\newenvironment{emp}{\pagebreak[1]\begin{empttn}\rm}{\end{empttn}}

\newtheorem{examfoo}[theoremfoo]{Example}
\newenvironment{example}{\pagebreak[1]\begin{examfoo}\rm}{\end{examfoo}}

\newtheorem{dfntn}[theoremfoo]{Definition}
\newenvironment{definition}{\pagebreak[1]\begin{dfntn}\rm}{\end{dfntn}}

\newtheorem{propositionfoo}[theoremfoo]{Proposition}
\newenvironment{proposition}{\pagebreak[1]\begin{propositionfoo}}{\end{propositionfoo}}
\newenvironment{prop}{\pagebreak[1]\begin{propositionfoo}}{\end{propositionfoo}}

\newenvironment{proof}
    {\pagebreak[1]{\narrower\noindent {\bf Proof:\quad\nopagebreak}}}{\QED}
\newenvironment{sketch}
    {\pagebreak[1]{\narrower\noindent {\bf Proof sketch:\quad\nopagebreak}}}{\QED}
\newenvironment{comment}{\penalty -50 $(*$\nolinebreak\ }{\nolinebreak $*)$\linebreak[1]\ }

\newenvironment{algorithm}[1]{\bigskip\noindent ALGORITHM~#1\renewcommand{\theenumii}{\arabic{enumii}}\renewcommand{\labelenumii}{Step \theenumii :}\begin{enumerate}}{\end{enumerate}END OF ALGORITHM\bigskip}

\newenvironment{protocol}[1]{\bigskip\noindent PROTOCOL~#1\renewcommand{\theenumii}{\arabic{enumii}}\renewcommand{\labelenumii}{Step \theenumii :}\begin{enumerate}}{\end{enumerate}END OF PROTOCOL\bigskip}

\newenvironment{red}[1]{\noindent REDUCTION~#1\renewcommand{\theenumii}{\arabic{enumii}}\renewcommand{\labelenumii}{Step \theenumii :}\begin{enumerate}}{\end{enumerate}END OF REDUCTION}

\newenvironment{con}{\noindent CONSTRUCTION\renewcommand{\theenumii}{\arabic{enumii}}\renewcommand{\labelenumii}{Step \theenumii :}\begin{enumerate}}{\end{enumerate}END OF CONSTRUCTION}

\newenvironment{alg}[1]{\bigskip\noindent ALGORITHM~#1\renewcommand{\theenumii}{\arabic{enumii}}\renewcommand{\labelenumii}{Step \theenumii :}\begin{enumerate}}{\end{enumerate}END OF ALGORITHM\bigskip}

\newcommand{\yyskip}{\penalty-50\vskip 5pt plus 3pt minus 2pt}
\newcommand{\blackslug}{\hbox{\hskip 1pt
        \vrule width 4pt height 8pt depth 1.5pt\hskip 1pt}}
\newcommand{\QED}{{\penalty10000\parindent 0pt\penalty10000
        \hskip 8 pt\nolinebreak\blackslug\hfill\lower 8.5pt\null}
        \par\yyskip\pagebreak[1]}

\newcommand{\BBB}{{\penalty10000\parindent 0pt\penalty10000
        \hskip 8 pt\nolinebreak\hbox{\ }\hfill\lower 8.5pt\null}
        \par\yyskip\pagebreak[1]}

\newcommand{\PYI}{CCR-8958528}
\newtheorem{factfoo}[theoremfoo]{Fact}
\newenvironment{fact}{\pagebreak[1]\begin{factfoo}}{\end{factfoo}}
\newenvironment{acknowledgments}{\par\vskip 6pt\footnotesize Acknowledgments.}{\par}

\newcommand{\xbar}{{\overline{x}}}
\newcommand{\reals}{{\sf R}}
\newcommand{\nat}{{\sf N}}
\newcommand{\ceil}[1]{\left\lceil {#1}\right\rceil}
\newcommand{\floor}[1]{\left\lfloor{#1}\right\rfloor}
\newcommand{\bit}{\{0,1\}}
\newcommand{\bits}[1]{\{0,1\}^{{#1}}}
\newcommand{\eig}{eig}
\newcommand{\SUM}{\ \ \sum_{k=\max\{ 0,j+m-n \} }^{\min\{j,m\}}\ \ }
\newcommand{\SUMa}{\ \ \sum_{j=\max\{ 0,a+m-n \} }^{\min\{a,m\}}\ \ }
\newcommand{\HAM}{{\rm HAM}}
\newcommand{\HAMa}{HAM_n^{(a)}}
\newcommand{\HAMasec}{HAM_{\lowercase{n}}^{({\lowercase{a}})}}
\newcommand{\HAMea}{HAM_n^{(=a)}}
\newcommand{\HAMeasec}{HAM_{\lowercase{n}}^{(=\lowercase{a})}}
\newcommand{\HAMeap}{HAM_n^{(=a')}}
\newcommand{\HAMeapp}{HAM_n^{(=n - a')}}

\centerline{\bf Lower bounds on the Deterministic and Quantum Communication Complexity}

\centerline{\bf of Hamming-Distance Problems}

\centerline{\bf by Andris Ambainis, William Gasarch, Aravind Srinivasan, Andrey Utis}

\begin{abstract}
Alice and Bob want to know if two strings of length $n$ are
almost equal. That is, do the strings differ on \textit{at most} $a$ bits?
Let $0\le a\le n-1$.
We show (1) any deterministic protocol -- as well as any
error-free quantum protocol ($C^*$ version) -- for this problem
requires at least $n-2$ bits of communication, and
(2) a lower bound of $n/2-1$ for error-free $Q^*$ quantum protocols.
We also show the same results for 
determining if two strings differ in \textit{exactly} $a$ bits.
Our results are obtained by lower-bounding the ranks of the
appropriate matrices. 
\end{abstract}

\section{Introduction}

Given $x,y\in \bits n$ one way to measure how much they
differ is the Hamming distance.

\begin{definition}
If $x,y\in \bits n$ then
$\HAM(x,y)$ is the number of bits on which $x$ and $y$ differ.
\end{definition}

If Alice has $x$ and Bob has $y$ then how many bits
do they need to communicate such that they both know $\HAM(x,y)$?
The trivial algorithm is to have Alice send $x$ (which takes $n$ bits)
and have Bob send $\HAM(x,y)$  (which takes $\ceil{\lg (n+1)}$ bits)
back to Alice. This takes $n+\ceil{\lg (n+1)}$ bits.
Pang and El Gamal~\cite{pang}
showed that this is essentially optimal.
In particular they showed that $\HAM$ requires 
at least $n +\lg(n+1-\sqrt n)$ bits to be communicated.
(See~\cite{abdel,commdoc,metzner,orlit} for more on
the communication complexity of $\HAM$.
See~\cite{securemulti} for how Alice and Bob can approximate
$\HAM$ without giving away too much information.)

What if Alice and Bob just want to know if $\HAM(x,y)\le a$?

\begin{definition}
Let $n\in\nat$. Let $a$ be such that $0\le a\le n-1$.
$\HAMa:\bits n \times \bits n \rightarrow \bit$ is the function

\begin{equation}
\HAMa(x,y) = 
\begin{cases}
1 & \hbox{ if $\HAM(x,y)\le a$;}\cr
0 &  \hbox{otherwise.}\cr
\end{cases}
\end{equation}
\end{definition}

The communication 
complexity of $\HAMa$ has been studied in various randomized 
and quantum settings by
Yao~\cite{qfingerprinting}, Gavinsky et al.~\cite{qpublic} (Section 6),
Gavinsky et al.~\cite{qfinger} (Section 3.2), 
and Huang et al.~\cite{Hamcomm}. 

How much communication is needed for this problem in the
deterministic model?
There is the trivial $(n+1)$-bit upper bound.
There is an easy reduction from equality on $n-a$ bits to $\HAMa$,
hence there is an easy $(n-a)$ lower bound.
In this paper we improve the lower bound. Note that this amounts
to improving the additive term.

We show the following:

\begin{enumerate}
\item\label{det:n-2}
For any $0 \leq a \leq n-1$, $\HAMa$ requires at least $n-2$ bits
in the deterministic model.
\item\label{det:n}
For $a\le  \frac{\sqrt{n}}{4}$, $\HAMa$ requires at least $n$ bits 
in the deterministic model.
\item\label{quant:n-2}
For any $0 \leq a \leq n-1$, $\HAMa$ requires at least $n-2$ bits
in the quantum model where Alice and Bob share an infinite number of EPR pairs,
using a classical channel, and always obtain the correct answer.
\item
For $a\le \frac{\sqrt{n}}{4}$, $\HAMa$ requires at least $n$ bits 
in the quantum model in item \ref{quant:n-2}.
\item\label{quant:n/2minus1}
For any $0 \leq a \leq n-1$, $\HAMa$ requires at least $\frac{n}{2}-1$ bits
in the quantum model where Alice and Bob share an infinite number of EPR pairs,
using a quantum channel, and always obtain the correct answer.
\item\label{quant:n/2}
For $a\le \frac{\sqrt{n}}{4}$, $\HAMa$ requires at least $\frac{n}{2}$ bits 
in the quantum model in item \ref{quant:n/2minus1}.
\end{enumerate}

Note that if $a=n$ then $(\forall x,y)[\HAMa(x,y)=1]$,
hence we do not include that case.

What if Alice and Bob need to determine if $\HAM(x,y)=a$ or not?

\begin{definition}
Let $n\in\nat$. Let $a$ be such that $0\le a\le n$.
$\HAMea:\bits n \times \bits n \rightarrow \bit$ is the function
\begin{equation}
\HAMea(x,y) =
\begin{cases} 
1 & \hbox{if $\HAM(x,y)=a$;}\cr
0 & \hbox{otherwise.}\cr
\end{cases}
\end{equation}
\end{definition}

We show the exact same results for $\HAMea$ as we do for $\HAMa$.
There is one minor difference: for $\HAMa$ the $a=n$ case had complexity 0
since all pairs of strings differ on at most $n$ bits;
however, for $\HAMea$ the $a=n$ case has complexity $n+1$ as it is
equivalent to equality.

All our results use the known ``log rank'' lower bounds on
classical and quantum communication complexity: Lemmas~\ref{le:rank}
and \ref{le:qrank}. Our approach is to lower-bound the ranks of the
appropriate matrices, and then to invoke these known lower bounds.

\section{Definitions, Notations, and Useful Lemmas}

We give brief definitions of both classical and quantum communication
complexity. See~\cite{commcomp} for more details on classical, 
and~\cite{qsurvey} for more details on quantum.

\begin{definition}
Let $f$ be any function from $\bits n \times \bits n$ to $\bit $.
\begin{enumerate}
\item
A \textit{protocol } for computing $f(x,y)$, where Alice has $x$ and Bob has $y$,
is defined in the usual way (formally using decision trees).
At the end of the protocol both Alice and Bob know $f(x,y)$.
\item
$D(f)$ is the number of bits transmitted in the optimal deterministic protocol for $f$.
\item
$Q^*(f)$ is the number of bits transmitted in the optimal quantum protocol where
we allow Alice and Bob to share an infinite number of EPR pairs
and communicate over a quantum channel. For quantum protocols, we fix the number of
qubits communicated in each round (assuming that in the first round Alice always communicates $c_1$ qubits, in the second round Bob communicates $c_2$ qubits and so on, where $c_1, c_2, \ldots$ are independent of inputs $x$ and $y$). 
\item
$C^*(f)$ is the number of bits transmitted in the optimal quantum protocol where
we allow Alice and Bob to share an infinite number of EPR pairs
and communicate over a classical channel.
\item
$M_f$ is the $2^n \times 2^n$ matrix where the rows and columns are indexed
by $\bits n$ and the $(x,y)$-entry is $f(x,y)$.
\end{enumerate}
\end{definition}

Let $\lg$ denote the logarithm to the base two.
Also, as usual, if $x < y$, then $\binom{x}{y}$ is taken to be zero. 

The following theorem is due to
Mehlhorn and Schmidt~\cite{ranklower}; 
see also \cite{commcomp}.

\begin{lemma}\label{le:rank}
If $f:\bits n \times \bits n \rightarrow \bit $
then
$D(f)\ge \lg({\rm rank}(M_f))$.
\end{lemma}

Buhrman and de Wolf~\cite{qlogrank} proved a 
similar theorem for quantum communication complexity.

\begin{lemma}\label{le:qrank}
If $f:\bits n \times \bits n \rightarrow \bit $
then the following hold.
\begin{enumerate}
\item
$Q^*(f)\ge \frac{1}{2}\lg({\rm rank}(M_f))$.
\item
$C^*(f)\ge \lg({\rm rank}(M_f))$.
\end{enumerate}
\end{lemma}

We will need the following definition and notation

\begin{definition}
The {\it Krawtchouk Polynomials} (see~\cite{kpoly} and the references therein) are polynomials that
are parameterized by $a,n,q\in\nat$ with $q$ a prime power
and are defined by
$$k_a(n,q;x) = \sum_{k=0}^a (-1)^k(q-1)^{a-k}\binom{x}{k}\binom{n-x}{a-k}$$
(In the paper~\cite{kpoly} they use $N$ instead of $n$ and order the variables as $k_a(x,q,N)$.)
\end{definition}

\begin{definition}
Let 
$$F(a,n;x)= \sum_{j=0}^a \sum_{k=\max\{0,j+x-n\}}^{\min\{j,x\}}  \binom{x}{k} \binom{n-x}{j-k} (-1)^k.$$
\end{definition}

\begin{definition}
$$G(a,n;x)= \sum_{k=\max\{0,a+x-n\}}^{\min\{a,x\}}\binom{x}{k}\binom{n-x}{a-k} (-1)^k=\sum_{k=0}^{a} \binom{x}{k} \binom{n-x}{a-k} (-1)^k.$$
(The equality comes from our convention:
if $a < b$, then $\binom{a}{b}$ is taken to be zero.)
Note that $G(a,n;x)=k_a(n,2;x)$.
\end{definition}

\subsection{Lemmas Useful for the Complexity of $\HAMa$}\label{se:HAMa}

\begin{definition}
Let $M_a$ be $M_{\HAMa}$, the $2^n\times 2^n$ matrix representing $\HAMa$.
\end{definition}

\begin{lemma}
$M_a$ has $2^n$ orthogonal eigenvectors.
\end{lemma}

\begin{proof}
This follows from $M_a$ being symmetric.
\end{proof}

We know that $M_a$ has $2^n$ eigenvalues; however, some of them may be 0.
We prove that $M_a$ has few 0-eigenvalues.
This leads to a lower bound on $D(\HAMa)$ by Lemma~\ref{le:rank}.

\begin{definition}\label{de:vz}
Let $z\in \bits n$.
\begin{enumerate}
\item
$v_z \in \reals^{2^n}$ is defined by, for all $x\in \bits n$,  $v_z(x) = (-1)^{\sum_i x_i z_i }$. The entries $v_z(x)$ of $v_z$ are ordered in the natural
way: in the same order as the order of the index $x$
in the rows (and columns) of $M_a$. 
\item
We show that $v_z$ is an eigenvector of $M_a$. Once that is done
we let $\eig(z)$ be the eigenvalue of $M_a$ associated with $v_z$.
\end{enumerate}
\end{definition}

\begin{lemma}\label{le:main}~
\begin{enumerate}
\item The vectors $\{v_z: ~z\in \bits n\}$ are orthogonal.
\item
For all $z\in \bits n$, $v_z$ is an eigenvector of $M_a$.
\item
If $z$ has exactly $m$ 1's in it,
then $eig(z)=F(a,n;m)$
\end{enumerate}
\end{lemma}

\begin{proof}
The first assertion (orthogonality) follows by simple counting.
We now prove the final two assertions together.
Let $z\in \bits n$ have exactly $m$ ones in it.

Fix a row in $M_a$ that is indexed by $x\in \bits n$.
Denote this row by $R_x$.
We need the following notation:
\[
\begin{array}{rl}
L_a = & \{ y \mid \HAM(x,y) \le a \}\cr
E_j = & \{ y \mid \HAM(x,y) = j \}\cr
\end{array}
\]
We will show that $R_x \cdot v_z$ 
is a constant multiple (independent of $x$) times $v_z(x)$.
Now,
$$R_x\cdot v_z =  \sum_{y\in \bits n} \HAMa(x,y) v_z(y) = \sum_{y\in L_a} v_z(y) = \sum_{y\in L_a}  (-1)^{\sum_i y_iz_i }.$$
We would like this to equal $b \times v_z(x)$ for some constant $b$.
We set it equal to $b\times v_z(x)$ and deduce which $b$'s work.
Suppose
$$b \times v_z(x) = \sum_{y\in L_a}  (-1)^{\sum_i y_iz_i }.$$

We have
\begin{eqnarray}
b & = & \frac{1}{v_z(x)} \sum_{y\in L_a}  (-1)^{\sum_i y_iz_i } \nonumber \\
& = & v_z(x) \sum_{y\in L_a}  (-1)^{\sum_i y_iz_i} \nonumber \\
& = & (-1)^{\sum_i x_iz_i} \sum_{y\in L_a}  (-1)^{\sum_i y_iz_i } \hbox{\ \ \ (by the definition of $v_z(x)$)} \nonumber \\
& = & \sum_{y\in L_a}  (-1)^{\sum_i (x_i+y_i)z_i} \nonumber \\
& = & \sum_{y\in L_a}  (-1)^{\sum_i |x_i-y_i|z_i } \hbox{\ \ \  (since $x_i+y_i \equiv |x_i-y_i| \pmod 2$) } \nonumber \\
& = & \sum_{j=0}^a \sum_{y\in E_j}  (-1)^{\sum_i |x_i-y_i|z_i } \hbox{\ \ \ (since $L_a = \bigcup_{j=0}^a E_j$)}. \label{eqn:b}
\end{eqnarray}

We partition $E_j$.
If $y\in E_j$ then
$x$ and $y$ differ in exactly $j$ places.
Some of those places $i$ are such that $z_i=1$.
Let $k$ be such that the number of places where $x_i\ne y_i$ and $z_i=1$.

\smallskip

\noindent
{\it Upper Bound on $k$:}
Since there are exactly $m$ places where $z_i=1$ we have $k\le m$.
Since there are exactly $j$ places where $x_i\ne y_i$ we have $k\le j$.
Hence $k\le\min\{j,m\}$.

\smallskip

\noindent
{\it Lower Bound on $k$:}
Since there are exactly $n-m$ places where $z_i=0$, we have $j-k\le n-m$.
Hence $k\ge \max\{0,j+m-n\}$.

\smallskip

In summary, the only relevant $k$ are
$\max\{0,j+m-n\} \le k \le \min\{j,m\}$.
Fix $j$. For 

\noindent
$\max\{0,j+m-n\}\le k\le \min\{j,m\}$, 
let $D_{j,k}$ be defined as follows:
$$D_{j,k} = \{ y \mid ((y \in E_j) \wedge 
(\hbox{on exactly $k$ of the coordinates 
where $x_i\ne y_i$, we have $z_i = 1$})) \}.$$

Note that

$$E_j = \bigcup_{k=0}^{\min\{j,m\}} D_{j,k}$$
and $|D_{j,k}|= \binom{m}{k} \binom{n-m}{j-k}$. So, by (\ref{eqn:b}),
$$
b = \sum_{j=0}^a \sum_{y\in E_j}  (-1)^{\sum_i |x_i-y_i|z_i }
  = \sum_{j=0}^a \SUM \sum_{y\in D_{j,k}} (-1)^{\sum_i |x_i-y_i|z_i }.
$$

By the definition of $D_{j,k}$ we know that for exactly $k$ of the values of $i$
we have both $|x_i-y_i|=1$ and $z_i=1$. On all other values one of the two
quantities is 0. Hence we have the following:
\begin{eqnarray*}
b & = & \sum_{j=0}^a \SUM  \sum_{y\in D_{j,k}} (-1)^k \\
& = & \sum_{j=0}^a \SUM  |D_{j,k}| (-1)^k \\
& = & \sum_{j=0}^a \SUM  \binom{m}{k}\binom{n-m}{j-k} (-1)^k.
\end{eqnarray*}

Notice that $b$ is independent of $x$ and is of the form required.
\end{proof}

\begin{definition}
Let 
$$F(a,n;m)= \sum_{j=0}^a \SUM \binom{m}{k}\binom{n-m}{j-k} (-1)^k.$$
\end{definition}

\begin{lemma}\label{le:uslea}~
\begin{enumerate}
\item
$D(\HAMa) \ge \lg \sum_{m: F(a,n;m)\ne 0 } \binom{n}{m}.$
\item
$Q^*(\HAMa) \ge \frac{1}{2}\lg \sum_{m: F(a,n;m)\ne 0 } \binom{n}{m}.$
\item
$C^*(\HAMa) \ge \lg \sum_{m: F(a,n;m)\ne 0 }\binom{n}{m}.$
\end{enumerate}
\end{lemma}

\begin{proof}
By Lemma~\ref{le:main}, the eigenvector $v_z$ has a nonzero eigenvalue
if $v_z$ has $m$ 1's and $F(a,n;m)\ne0$.
The rank of $M_a$ is the number of nonzero eigenvalues that correspond
to linearly independent eigenvectors. 
This is $\sum_{m: F(a,n;m)\ne 0 }\binom{n}{m}.$
The theorem follows from Lemmas~\ref{le:rank} and \ref{le:qrank}.
\end{proof}

\begin{lemma}\label{le:lea}
The number of values of $m$ for which $F(a,n;m)=0$ is $\le a$.
\end{lemma}

\begin{proof}
View  the double summation $F(a,n;m)$ as a  polynomial in $m$. 
We first show that $F(a,n;m)$ is not identically zero.
Plug in $m=n$.  Then

$$F(a,n;n) = \sum_{j=0}^a \sum_{k=\max\{0,j\} }^{\min\{j,n\}} \binom{n}{k}\binom{0}{j-k} (-1)^k=
\sum_{j=0}^a \sum_{k=j}^{j} \binom{n}{k}\binom{0}{j-k} (-1)^k=
\sum_{j=0}^a \binom{n}{j}\binom{0}{0} (-1)^j
$$
Since $0\le a<n$ this cannot be 0.

We now show that $F(a,n;m)$ has degree $a$ and hence has at most $a$ roots.
The $j$th summand has degree $k+(j-k)=j$.
Since $j\le a$ the entire sum can be written as a polynomial in
$m$ of degree $a$. This has at most $a$ roots.
\end{proof}

\subsection{Lemmas Useful for the Complexity of $\HAMea$}

\begin{definition}
Let $M_{=a}$ be $M_{\HAMea}$, the $2^n\times 2^n$ matrix representing $\HAMea$.
\end{definition}

The vectors $v_z$ are the same ones defined in Definition~\ref{de:vz}.
We show that $v_z$ is an eigenvector of $M$. Once that is done
we let $\eig(z)$ be the eigenvalue of $M_{=a}$ associated to $z$.

The lemmas needed, and the final theorem,
are very similar (in fact easier) to those
in the Section~\ref{se:HAMa}. Hence we just
state the needed lemmas and final theorem.

\begin{lemma}\label{le:maina}~
\begin{enumerate}
\item
For all $z\in \bits n$ $v_z$ is an eigenvector of $M_{=a}$.
\item
If $z$ has exactly $m$ 1's in it then $eig(z)=G(a,m;n)$.  
\end{enumerate}
\end{lemma}

\begin{lemma}\label{le:useqa}~
\begin{enumerate}
\item
$D(\HAMea) \ge \lg \sum_{m: G(a,n;m)\ne 0 }\binom{n}{m}.$
\item
$Q^*(\HAMea) \ge \frac{1}{2} \lg \sum_{m: G(a,n;m)\ne 0 }\binom{n}{m}.$
\item
$C^*(\HAMea) \ge \lg \sum_{m: G(a,n;m)\ne 0 }\binom{n}{m}.$
\end{enumerate}
\end{lemma}

\section{The Complexity of $\HAMasec$ and $\HAMeasec$ for $\lowercase{a}\le \frac{\sqrt{\lowercase{n}}}{4}$}\label{se:hamasq}

\begin{theorem}\label{th:main}
If $a\le \frac{\sqrt n}{4}$
then the following hold.
\begin{enumerate}
\item
$D(\HAMa) \ge n$.
\item
$Q^*(\HAMa) \ge n/2$.
\item
$C^*(\HAMa)\ge n$.
\end{enumerate}
\end{theorem}

\begin{proof}
By Lemma~\ref{le:uslea}
$D(f),Q^*(f) \ge \lg (\sum_{m: F(a,n;m)\ne 0 }\binom{n}{m})$ and
$C^*(f) \ge \frac{1}{2}\lg (\sum_{m: F(a,n;m)\ne 0 }\binom{n}{m})$.

Note that
$$2^n = \sum_{m: F(a,n;m)\ne 0 }\binom{n}{m}+\sum_{m: F(a,n;m)=0 }\binom{n}{m}.$$
By Lemma~\ref{le:lea} $|\{ m : F(a,n;m)=0 \}|\le a$. Hence,
$$\sum_{m: F(a,n;m)=0 } \binom{n}{m} \le |\{ m : F(a,n;m)=0 \}|\cdot\max_{0\le m\le n}\binom{n}{m}
\le a \binom{n}{n/2} \le \frac{a2^n}{\sqrt n}.$$
So, if $a\le \frac{1}{4}\sqrt n$, then 
$$\sum_{m: F(a,n;m)\ne 0 }\binom{n}{m}\ge 2^n - \frac{a2^n}{\sqrt n}
\geq 2^n - 2^{n-2}.$$
Hence,
$$\lg \left(\sum_{m: F(a,n;m)\ne 0 }\binom{n}{m}\right) \ge 
\lg(2^n - 2^{n-2}); ~~
\mathrm{i.e.}, ~
\left\lceil \lg \left(\sum_{m: F(a,n;m)\ne 0 }\binom{n}{m}\right) 
\right\rceil \ge n.$$
Therefore we have out lower bounds.
\end{proof}

The following theorem has a proof that is very similar to the proof of Theorem~\ref{th:main};
hence we omit it.

\begin{theorem}\label{th:maine}
If $a\le \frac{\sqrt n}{4}$ then the following hold.
\begin{enumerate}
\item
$D(\HAMea) \ge n$.
\item
$Q^*(\HAMea) \ge n/2$.
\item
$C^*(\HAMea)\ge n$.
\end{enumerate}
\end{theorem}

\section{The Complexity of $\HAMasec$ and $\HAMeasec$ for General $\lowercase{a}$}\label{se:gen}

Recall that $G(a,n;x)$ is the  Krawtchouk polynomial $k_a(n,2;x)$.

\begin{lemma}\label{le:kzero}
For all $a,n$ let 
$r_{a,1}^n < r_{a,2}^n < \cdots < r_{a,a}^n$ be the  roots of the poly $k_a(n,q;x)$.
(They need not be integers.)
\begin{enumerate}
\item
For all $i$ there is an integer in the open interval $(r_{a,i}^n,r_{a,i+1}^n)$.
\item
Let $m$ be an integer.
If $k_a(n,q;m)=0$ then $k_a(n,q;m+1)\ne 0$.
\item
Let $m$ be an integer.
If $G(a,n;m)=0$ then $G(a,n;m+1)\ne 0$.
\end{enumerate}
\end{lemma}

\begin{proof}

\noindent
1) This is from~\cite{orthpoly}.

\noindent
2) Assume, by way of contradiction, that there is an integer such that $k_a(n,q;m)=0$ and $k_a(n,q;m+1)=0$.
By part 1 there is an integer in the open interval $(m,m+1)$. This is a contradiction.

\noindent
3) This follows from the fact that $G(a,n;x)=k_a(n,2;x)$.
\end{proof}

\begin{theorem}
\label{thm:eqcomplexity}
For large enough $n$ and all $0 \leq a \leq n$ the following hold.
\begin{enumerate}
\item
$D(\HAMea) \geq n-2$.
\item
$Q^*(\HAMea)\geq \frac{n}{2}-1$.
\item
$C^*(\HAMea)\ge n-2$.
\end{enumerate}
\end{theorem}

\begin{proof}
First suppose $a \leq n/2$. Note that
\begin{equation}
\label{eqn:small-a-large-m}
\sum_{m: G(a,n;m)\ne 0}\binom{n}{m} \geq
\sum_{m \geq n/2: G(a,n;m)\ne 0}\binom{n}{m}.
\end{equation}

Lemma~\ref{le:kzero} shows that no two consecutive values of $m$ in 
the range $a\le m\le n$ (and hence in the range $n/2 \le m\le n$) satisfy the
condition ``$G(a,n;m) = 0$''.
Hence our problem is to minimize the sum of a subset of 
$$\left\{\binom{n}{n/2},\binom{n}{n/2-1},\ldots,\binom{n}{0} \right\},$$
where if we omit $\binom{n}{i}$, we must use $\binom{n}{i-1}$.
Since $\binom{n}{m}$ decreases in the range $n/2\le m\le n$, this sum is minimized by taking every other term: thus this sum is
always at least $2^{n-2}$. Our theorem follows from Lemma~\ref{le:useqa}.

Now we apply symmetry to the case $a> n/2$: note that Alice can
reduce the problem with parameter $a$ to the problem with parameter
$n - a$, simply by complementing each bit of her input $x$. Thus,
the same communication complexity results hold for the case $a > n/2$.
\end{proof}

\begin{lemma}\label{le:zero}
Let $0 \leq a < m < n$, and suppose $F(a,n;m)=0$. Then $F(a,m+1;n) \neq 0$.
\end{lemma}

\begin{proof}
We will use the terminology and methods of generating functions.

\noindent
{\bf Notation}
$[x^b]g(x)$ is the coefficient of $x^b$ in the power series
 expansion of $g(x)$ around $x_0=0$.

\begin{lemma}\label{le:easy}~
\begin{enumerate}
\item
If $a\in\nat$ and $f(x)$ is any power series then
$$\sum_{j=0}^a [x^j]f(x) = [x^a](f(x)\sum_{j=0}^\infty x^j) = [x^a]\frac{f(x)}{1-x}.$$
\item
$$G(j,n;m)=(-1)^m[x^j]((x-1)^m(x+1)^{n-m}).$$
\item
$$F(a,n;m)=\sum_{j=0}^a G(j,n;m).$$
\end{enumerate}
\end{lemma}

\begin{proof}
Items 1 and 3 are clear. We prove item 2.
We show $G(a,n;m)=(-1)^m[x^a]((x-1)^m(x+1)^{n-m})$ for ease of notation;
however, the proof clearly holds for $j$ instead of $a$.

$$(x-1)^m (x+1)^{n-m} = \sum_{k=0}^m \binom{m}{k} x^k (-1)^{m-k}\sum_{j=0}^{n-m} \binom{n-m}{j} x^j$$

$$(x-1)^m (x+1)^{n-m} = \sum_{k=0}^m\sum_{j=0}^{n-m} \binom{m}{k} x^k (-1)^{m-k}\binom{n-m}{j} x^j$$

$$(x-1)^m (x+1)^{n-m} = \sum_{k=0}^m\sum_{j=0}^{n-m} \binom{m}{k} \binom{n-m}{j} x^{k+j} (-1)^{m-k} $$
 
The coefficient of $x^a$ is

$\sum_{k=0}^a \binom{m}{k} \binom{n-m}{a-k} (-1)^{m-k}$ which is $(-1)^mG(a,n;m)$.
\end{proof}

Using Lemma~\ref{le:easy} we obtain the following.

\[
\begin{array}{rl}
 F(a,n;m)=&\sum_{j=0}^a G(j,n;m)=(-1)^m \sum_{j=0}^a [x^j]((x-1)^m(x+1)^{n-m})\cr
         =&(-1)^m [x^a]((x-1)^m(x+1)^{n-m}\cdot\frac{1}{1-x})\cr
         =&(-1)^{m-1} [x^a]((x-1)^{m-1}(x+1)^{n-m})=G(a,n-1;m-1).\cr
\end{array}
\]

 Hence $F(a,n;m)=F(a,n;m+1)=0$ iff
 $G(a,n-1;m-1)=G(a,n-1;m)=0$. But the latter is impossible
 by Lemma~\ref{le:kzero}, thus the lemma is proved.
\end{proof}

\begin{theorem}
For large enough $n$ and all $0 \leq a \leq n-1$, 
the following hold.
\begin{enumerate}
\item
$D(\HAMa) \geq n-2$.
\item
$Q^*(\HAMa) \geq \frac{n}{2}-1$.
\item
$C^*(\HAMa) \geq n-2$.
\end{enumerate}
\end{theorem}

\begin{proof}
The proof is identical to that of Theorem~\ref{thm:eqcomplexity}
except for one point. In that proof we obtained the $a> n/2$
case easily from the $a\le n/2$ case. Here it is also easy
but needs a different proof. Let $a> n/2$ and, for all $x\in \bits n$,
let $\xbar$ be obtained from $x$ by flipping every single bit.
Note that

$\HAMa(x,y)=1$ iff $\HAM(x,y)\le a$ iff $\HAM(\xbar,y)\ge n-a$
iff NOT($\HAM(\xbar,y)\le (n-a)-1$ iff
$\HAM_n^{n-a-1}(\xbar,y)=1$.

Since $n-a-1 \le n/2$ we have that a lower bound for the $a\le n/2$
case implies a lower bound for the $a>n/2$ case.
\end{proof}

\section{Open Problems}

We make the following conjectures.

\begin{enumerate}
\item
For all $n$, for all $a$, $0\le a\le n-1$, $D(\HAMa)=C^*(\HAMa)=n+1$ 
\item
For all $n$, for all $a$, $0\le a\le n-1$, $Q^*(\HAMa)=\frac{n}{2}+1$.
\item
For all $n$, for all $a$, $0\le a\le n$, $D(\HAMea)=C^*(\HAMea)=n+1$.
\item
For all $n$, for all $a$, $0\le a\le n-1$, $Q^*(\HAMea)=\frac{n}{2}+1$.
\end{enumerate}

The first and third conjecture are just a matter of improving the lower
bound by 3 bits. For the second and fourth conjecture, superdense coding~\cite{superdense} provides an upper
bound of $\frac{n}{2}+1$ on $Q^*(\HAMa)$ and on $Q^*(\HAMea)$ ($\frac{n}{2}$ qubits for Alice to communicate her input $x$ to Bob and 1 bit for Bob to communicate the function value $f(x, y)$ back to Alice). The remaining part is to improve the lower bound by 2 qubits.

\section{Acknowledgement}
An earlier version of this work appeared in the \emph{Proc.\ International Symposium on Algorithms and Computation} (ISAAC), 2006; 
We would like to thank an anonymous referee of the conference version who pointed
out the connection to Krawtchouk polynomials. We also thank the journal referees for their helpful suggestions. 
This work is supported by the National Science Foundation, under
grants CCR-01-05413, CCR-02-08005, CCF 14-22569, CNS-1010789, and CCF-1422569.

%\bibliographystyle{abbrv}
%\bibliography{bibfile}

\end{document}